\documentclass[aps,prb,twocolumn,showpacs]{revtex4-1}

\usepackage{graphicx} % Include figure files
\usepackage{amssymb}
\usepackage{color}
\usepackage[caption=false, ...]{subfig}
\usepackage{amsmath}
\usepackage{hyphenat}
\usepackage{hyperref}

\begin{document}
\hyphenpenalty 10000

\title{Quasi-particle band structures and optical properties of the ``magic sequence'' SiGe superstructure}

\author{Mohammad Reza Ahmadpour Monazam}\email{Mohammad\_Reza.Ahmadpour\_Monazam@jku.at}
\author{Kurt Hingerl}\email{Hingerl.Kurt@jku.at}
\affiliation{Zentrum f\"ur Oberfl\"achen- und Nanoanalytik, Altenberger Str.69, 4040 Linz, Austria}

\author{Peter Puschnig}\email{peter.puschnig@uni-graz.at}
\affiliation{Institut f\"ur Physik, Karl-Franzens-Universit\"at Graz, Universit\"atsplatz 5, 8010 Graz, Austria}

\date{\today}

\begin{abstract}
The quasi-particle band structure and dielectric function for the so-called magic sequence SiGe$_2$Si$_2$Ge$_2$SiGe$_{12}$ (or $\alpha_{12}$) structure [PRL 108, 027401 (2012)] are calculated by many-body perturbation theory (MBPT) within an ab-initio framework. On top of density functional calculations within the local density approximation (LDA) leading to a fundamental band gap of 0.23 eV, we have computed the quasi-particle band structure within the G$_0$W$_0$ approach opening the gap to 0.61 eV. Moreover, we have calculated the optical properties by solving the Bethe-Salpeter equation (BSE) for the electron-hole two-particle correlation function. When comparing the imaginary part of the dielectric function obtained at various levels of approximation, \emph{i.e.} the independent particle approximation (or random phase approximation) based on (i) the LDA or (ii) GW band structures, and (iii) the BSE including local field effects and electron-hole correlations, we observe that the important first transition is better explained with taking into account excitonic effects. Moreover, the onset transition originating from the direct transition of the magic sequence structure is also investigated.       
\end{abstract}
\maketitle

%%%%%%%%%%%%%%%%%%%%%%%%%%%%%%%%%%%%%%%%%%%%%%%% Introduction %%%%%%%%%%%%%%%%%%%%%%%%%%%%%%%
\section{Introduction}
Optoelectronic industry is mainly focused on group \textit{III-V} and \textit{II-VI} alloys to design optically active semiconductors at the desired 1.55 $\mu$m wavelength for transmitting signals.\cite{Desurvire1994} However, such compounds have a considerably large lattice mismatch with silicon which impedes their incorporation with current high-speed electronic technology based on Si. On the other hand, the use of  group \textit{IV} semi-conductors, and in particular Si, as optically active materials is strongly limited  due to their indirect fundamental bandgap nature, making them poor emitters of light. Thus, much attention has been paid over the last decades to grow lattice matched materials exhibiting efficient light emission and absorption that could be integrated on Si in single chip as electronic-optical component.\cite{Froyen1987,Satpathy1988,M.IkedaK.Terakura1993}

Here, Si-Ge superlattice structures have attracted a lot of interest due to their close lattice match with silicon substrates. High-quality Si-Ge layered structures with perfect interfaces have been shown to result in better optoelectronic efficiencies than bulk Si-Ge alloys. \cite{Smith1990} Different multiples of (Si)$_m$/(Ge)$_n$ superlattices in various strain regimes were the subject of numerous papers over the past decades.\cite{Green2001,Zhang2001,Busby2010} Driven by the needs from application in semiconductor industry, the quest was to find a SiGe superstructure with the two distinctive properties: first, a direct band gap and, second, a strongly allowed optical transition. There is consensus in the literature that strain engineering is necessary for obtaining direct bandgap materials with these superlattices and that direct bandgap materials with these superlattices could not be feasible by using Si substrates. Instead, either germanium or alloyed Si-Ge substrates are used. Calculations and experiments have shown that prototypes of ten layered superlattices of Si/Ge have come close to direct bandgap characteristics and they also enhance optical transition, such as (Si)$_4$/(Ge)$_6$. Although there is a discrepancy between the lowest transition energy obtained by \textit{ab initio} calculations and experiment,\cite{Pearsall1989,Zachai1990,Schmid1991} there is agreement that the optical dipole moment for this transition is still small.

\begin{figure}[!htb]
	\includegraphics[width=\columnwidth]{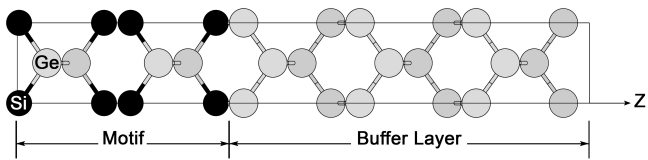}
	\caption{\label{structure}The layer sequence of the magic $\alpha_{12}$ structure consisting of the SiGe$_2$Si$_2$Ge$_2$Si motif and 12 Ge buffer layers.}
%  \label{structure}
\end{figure}

The recent proposal of a ``magic" Si-Ge stacking sequence discovered from an inverse engineering method by Manyeul d'Avezac and co-workers holds a promise for delivering a solution to this traditional semiconductor physics problem of finding a Si alloy with a direct and allowed band gap.\cite{Franceschetti1999,Avezac2012} Based on calculations within the framework of density functional theory (DFT), they predict that enhancements of the optical dipole moment up to 50 times with respect to the older (Si)$_4$/(Ge)$_6$ super-structure can be achieved. Their most promising structure consists of the motif SiGe$_2$Si$_2$Ge$_2$Si plus $n$ buffer layers of Ge, where $n=12-32$, which they denoted as ``$\alpha$" structure, Fig.~\ref{structure} shows this structure, with the motif consisting of Si and Ge atoms, the buffer layer only of Ge, the crystal growth direction is along $z$. Employing DFT calculations at the level of density functional approximation (LDA) shows that $\alpha_{12}$ exhibits both a direct bandgap and optically allowed transitions between valance band maximum (VBM) and conduction band minimum (CBM). They have also demonstrated the robustness of the $\alpha$ structure in the sense that small deviations from the ideal layer sequence will retain the direct band gap property. \cite{Avezac2012}

While the DFT calculations by d'Avezac {\em et al.} for the $\alpha$ structure clearly show an enhancement of the direct transition compared to previously investigated Si/Ge superlattices, DFT lacks rigorous justification when applied to excited state properties such as band structure or optical properties.
For instance, it is well-known that the band gap of semiconductors and consequently the onset of transitions in their dielectric function is largely underestimated due to missing derivative discontinuity in commonly used approximations for the exchange-correlation functional such as the LDA or the generalized gradient approximation (GGA). Moreover, correlations between the excited electron and the remaining hole taking place in optical transitions are also neglected when employing single-particle approximation methods. In order to overcome these shortcomings, the use of many-particle methods is inevitable to get a better and more reliable comprehension of electronic and optical mechanisms in such structures. Using many-body perturbation theory, it is possible to conceptually improve the first-principles description of the quasi-particle band structure and optical properties. By utilizing the so-called GW approximation for the electron self energy band gaps turn out to be in good agreement with experiment, and by solving the Bethe-Salpeter equation (BSE) for the electron-hole pair, electron-hole correlations, that is excitonic effects, can be included.\cite{Onida2002}

In this work, we use self-consistent density functional calculations within a norm-conserving pseudo-potential approach as starting point for subsequent G$_0$W$_0$ and BSE computations. Thereby, we achieve a greatly improved description of the  band gap compared to the DFT band structure.
By solving the Bethe-Salpeter equation, we obtain the frequency dependence of dielectric function including electron-hole correlations. By comparing optical properties within the independent particle approximation (RPA) and from the solution of the BSE, we analyze the impact of electron-hole correlations. We conclude that the first transition is better explained when taking into account excitonic effects and that the onset transition originating from the direct transition of the magic sequence structure is also modified due to excitonic effects. Our paper is organized as follows: In Section II, the theoretical methodology for GW and BSE calculations are briefly explained. The G$_0$W$_0$ band structure and also the frequency dependence of dielectric function from solving BSE are discussed in Section III and the last section is devoted to the conclusion. Convergence tests for both computational approaches used in the paper are summarized up in an Appendix.

%%%%%%%%%%%%%%%%%%%%%%%%%%%%%%%%%%%%%%%%%%%%%%%%%%% Computational Approach %%%%%%%%%%%%%%%%%%%%%%%%%%%

\section{Computational approach}

\subsection{Ground state calculations}
The starting point for the investigation are ground-state DFT calculations for which we employ the plane-wave code ABINIT.\cite{Gonze2002,Gonze2005} 
For all calculations, we consider the $\alpha_{12}$ structure composed of 20 atoms of silicon and germanium in sequence of SiGe$_{2}$Si$_{2}$Ge$_{2}$Si plus 12 layers of Ge atoms. In order to take into account the growth conditions on a Si$_{0.4}$Ge$_{0.6}$ substrate, we used the lattice parameter of Si$_{0.4}$Ge$_{0.6}$ with in-plane component as $a_{\parallel}=5.55$~{\AA}. By keeping the lateral lattice parameters fixed, the relaxation of the $c$-axis of the  $\alpha_{12}$ structure leads to a value of $a_{\perp}=28.07$~{\AA}. Fig.~\ref{structure} shows schematically the $\alpha_{12}$ structure which is used for all subsequent calculations.

We use norm-conserving Troullier-Martins scheme pseudo-potentials for the treatment of electron-ion interactions and approximate exchange-correlation effects within the local density approximation (LDA). A cut-off energy of 50 Ry is used for the expansion of wave functions in a plane wave basis set, and a  16$\times$16$\times$2 uniform k-point grid has been used for Brillouin zone (BZ) integrations.

\subsection{GW calculations}
The Kohn-Sham energies and orbitals build the basis for the subsequent quasi-particle calculations according to Hedin's $GW$ approximation.\cite{Onida2002,Hedin1965} Here, the self energy $\Sigma$ in frequency space is a convolution of the Green's function $G$, and  the screened Coulomb interaction $W$
\begin{equation}
\label{self_energy}
\Sigma^{GW}(r,r';\omega)=\frac{i}{2\pi}\int{e^{i\omega'\delta^{+}}G\left(r,r';\omega+\omega'\right)W\left(r,r';\omega'\right)d\omega'},
\end{equation}
where $\delta$ is a positive infinitesimal value. 
In the $GW$ approximation, vertex corrections are neglected in the calculation of the polarizability and the self-energy.  Furthermore, in the non-selfconsistent G$_0$W$_0$ approach,  one uses DFT orbitals and energies in order to construct the Greens function, $G_0$, and to set up the screened Coulomb interaction, $W_0$. Quasi-particle energies are then obtained by treating the difference between the self-energy and the Kohn-Sham exchange-correlation energy as perturbation, and one obtains from first-order perturbation theory:
\begin{equation}
\label{qp-energies}
\epsilon^{QP}=\epsilon^{KS}+Z\left<\psi^{KS}\left|\Sigma\left(\epsilon^{KS}\right)-v_{xc}\right|\psi^{KS}\right>
\end{equation}
where
\begin{equation}
 Z=\biggr[1-\left<\psi^{KS}\left|\frac{\partial\Sigma\left(\epsilon\right)}{\partial\epsilon^{KS}}\right|\psi^{KS}\right>\biggr]^{-1}
\end{equation}
is the renormalization factor. 

For our computations, we utilize the $GW$ implementation of ABINIT. \cite{Gonze2009} We calculate quasi-particle corrections on a 16$\times$16$\times$2 mesh in the Brillouin zone, and use plane wave cut-offs of 50 Ry for the wave functions and 6 Ry for dielectric matrix. A convergence study of the quasi-particle gap at $\Gamma$ with respect to the latter cut-off can be seen in Fig.~\ref{gap_ecuteps} of the Appendix. We employ the effective-energy technique (EET) to avoid costly summations over unoccupied states in calculating the  RPA polarizability and in the correlation part of the self-energy.\cite{Berger2010,Puschnig2012}
In this work, the frequency dependence of $W_0$ is described by the plasmon-pole model according to Godby and Needs for which the dielectric matrix is evaluated at two frequencies: in the static limit  and an imaginary frequency close to the plasmon frequency.\cite{Godby1989}

\subsection{BSE calculations}
For BSE calculations we also used the implementation in the ABINIT code and employ two commonly used simplifications. First, we utilize the Tamm-Dancoff approximation (TDA) in which only the resonant part of the BSE Hamiltonian is kept, while the coupling between the resonant and antiresonant part is disregarded.\cite{Onida2002} Second, we approximate the screened Coulomb interaction in the static limit thereby neglecting dynamical screening effects in the e-h interaction.\cite{Bechstedt1997} While going beyond the TDA has been shown to be significant for the electron energy loss spectra (EELS) of silicon\cite{Olevano01} and the exciton binding energies in organic molecules and carbon nanotubes,\cite{Gruning2009} the TDA is expected have a smaller influence on the optical absorption spectra of bulk-like inorganic semiconductors. Dynamical screening effects, on the other hand, can be also be safely neglected when excitonic binding energies are small compared to the plasma frequencies (or the quasiparticle gap) of the investigated system,\cite{Ma2009} This requirement is fulfilled for the system under study since -- in retrospect -- exciton binding energies are quite small.
Using these two simplifications, the electron-hole Hamiltonian is a Hermitian operator and the exciton energies are obtained as eigenvalues of an effective electron-hole ($e^{-}-h^{+}$) Hamiltonian,
\begin{equation} \label{Hamiltonian}
\sum_{v'c'\mathbf{k'}} H_{vc\mathbf{k},v'c'\mathbf{k'}}^{exc} A_{v'c'\mathbf{k'}}^{\lambda}=E^{\lambda}A_{vc\mathbf{k}}^{\lambda}
\end{equation}
Here, $E^{\lambda}$ denote the excitonic energies, and $A_{vck}^{\lambda}$ are coupling coefficients which are used to construct the $e^{-}-h^{+}$ wavefunctions from the single-particle valence and conduction band orbitals. The Hamiltonian has the explicit form
\begin{equation} \label{kernel}
H_{vc\mathbf{k},v'c'\mathbf{k'}}^{exc}=\left(E_{c\mathbf{k}}-E_{v\mathbf{k}}\right)\delta_{vv'}\delta_{cc'}\delta_{\mathbf{k}\mathbf{k'}}+\Xi_{vc\mathbf{k},v'c'\mathbf{k'}}
\end{equation}
where, $\Xi_{vck,v'c'k'}=\left[2\bar{v}_{\left(vck\right)\left(v'c'k'\right)}-W_{\left(vck\right)\left(v'c'k'\right)}\right]$ is the BSE kernel. Here, $\bar{v}$ and $W$ denote the  matrix elements of the exchange and screened Coulomb terms, respectively. The $v$($c$) indices refer to valance (conduction) bands. Using the eigenvalues $E^{\lambda}$ and eigenvectors $A_{vc\mathbf{k}}^{\lambda}$ of $H^{exc}$ together with the transition matrix elements $\langle{v\mathbf{k}}\left|e^{-i\mathbf{qr}}\right|c\mathbf{k}\rangle$ between valence and conduction band states, the imaginary part of macroscopic dielectric function with electron-hole correlation effects is expressed in the following form: 

\begin{equation}
\epsilon^{M}_2\left(\omega\right)={\displaystyle\lim_{\mathbf{q}\rightarrow{0}}}\frac{8\pi^2}{\Omega}{\frac{1}{\mathbf{q^2}}}\sum_{\lambda}{\left|\sum_{vc\mathbf{k}} A_{vc\mathbf{k}}^{\lambda}\langle{v\mathbf{k}}\left|e^{-i\mathbf{qr}}\right|c\mathbf{k}\rangle\right|}^2\times\delta\left(E^{\lambda}-\omega\right).
\end{equation}
However, instead of directly diagonalizing the Hamiltonian of Eq.~\ref{Hamiltonian}, which gets computationally expensive for a large number of $k$-points and bands in the transition space, we used an iterative technique, the so-called Haydock recursion method to construct basis sets in which the Hamiltonian is tridiagonal. The number of iterations is roughly independent of the problem size and the iterative scheme is also less memory demanding at the expense that individual transition energies are not accessible and an empirical lifetime broadening $\eta$ has to be included to enable convergence of the spectra. 

The macroscopic dielectric function is obtained at two different theoretical levels of approximation: First, by setting the kernel $\Xi$ of Eq.~(\ref{kernel}) zero, the random phase approximation (RPA) without local field effects is obtained. By considering the exchange term only, it is possible to get an estimate of local field effects on RPA results.
%Thus, the RPA neglects both the exchange and direct electron-hole interaction matrix elements. 
Second, by taking into account both terms in $\Xi$ we receive results including excitonic effects which we will label as ''BSE'' in the following. The BSE calculations, in turn, are performed at two different levels of sophistication. First, we approximate the quasi-particle band structure that enters Eq.~(\ref{Hamiltonian}) by the Kohn-Sham band structure corrected by a simple scissors operator (s.o.) shift, \emph{i.e.} a $k$-independent upward shift of the Kohn-Sham conduction states as depicted in Fig.~\ref{bandstructure}. This has the advantage that $G_0W_0$ corrections only have to be computed at a single $k$ point. In the following, these results are labeled as ''BSE(s.o.)''. Finally, we will also present results which are based on the $G_0W_0$ quasi-particle band structure computed on a dense $k$ grid in the full BZ which we will denote as ''BSE(qp)''.

When solving the BSE, a number of convergence parameters must be controlled. In the Appendix, we include detailed convergence studies with respect to the most important computational parameters. Here, we only summarize the set of parameters which have been deduced from these convergence checks and have been further utilized for the calculations presented in the next section. For representing wavefunctions, we used a plane wave cut-off of 50 Ry, whereas for the response functions cut-off, we find a value of 6 Ry to be sufficient. This leads to quasi-particle corrections that are accurate to within 0.1 eV and well-converged BSE spectra (see Fig.~\ref{bse_ecuteps}). For setting up the BSE Hamiltonian, a total number of 40 bands (20 number of bands for both valence and conduction bands)  are adequate for the transition space as can be seen from Fig.~\ref{conv_bands}. While the $k$ point density is less critical for obtaining converged quasi-particle corrections, optical spectra are known to be quite sensitive to the size of the $k$ grid density. We have used both $\Gamma$-centered $k$ grids as well as meshes shifted by symmetry-breaking translation of $\{0.11, 0.21, 0.31\}$ (in fractions of the grid spacing) that are supposed to have a better convergence at the same grid density compared to non-shifted ones. The typical convergence of BSE spectra can be seen from Figs.~\ref{epsilon2_qp_energy} and \ref{conv_kpoints} of the Appendix. We conclude that a 14$\times$14$\times$2 for the shifted mesh can be used, while a slightly denser mesh of 16$\times$16$\times$2 grid for the $\Gamma$-centered grid leads to  acceptably converged spectra. 

Note however that a BZ sampling including $\Gamma$ point is preferable in case information about the lowest optical transitions that is a direct transition at the $\Gamma$ point. As a technical remark we also note that with the current implementation of the $GW$+BSE package within ABINIT, spectra at the BSE(qp) were only possible with the $\Gamma$-centered grids. Finally, a 16$\times$16$\times$2 $k$-grid together with a transition space of 20 valence and 20 conduction bands leads to a Hamiltonian matrix size of roughly $200,000$.

%%%%%%%%%%%%%%%%%%%%%%%%%%%%%%%%%%%%%%%%%%%%%%%%%%%%%%%%%%    Results %%%%%%%%%%%%%%%%%%%%%%%%%%%%%%%%%

\section{Results and discussion}
\subsection{GW Results}

\begin{figure}[!htbp]
	\includegraphics[width=\columnwidth]{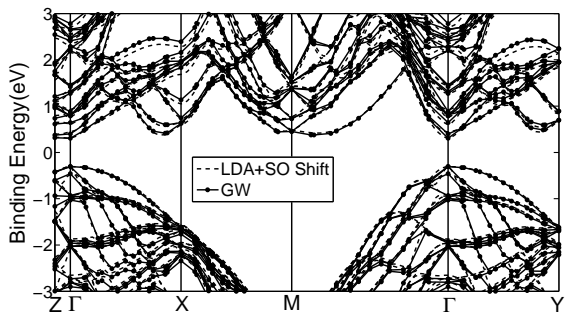}
        \caption{\label{bandstructure}The band structure of $\alpha_{12}$. The dashed line is LDA/DFT and dotted solid lines are G$_0$W$_0$ results. Note that the LDA CBs are shifted by 0.38 eV according the $G_0W_0$ calculation in order to facilitate a better comparison with the G$_0$W$_0$ band structure. }
%\label{bandstructure}
\end{figure}

The band structure of $\alpha_{12}$ is presented in Fig.~\ref{bandstructure} for both LDA and $GW$ calculations along high symmetry lines in the Brillouin zone. The dashed line shows the LDA results with exhibiting a direct band gap in agreement with Ref.~\onlinecite{Avezac2012}. Note that we have shifted the CBs by a rigid scissors shift of $0.38$ eV in order to facilitate a better comparison with the $GW$ band structure. The dotted solid lines are our $G_0W_0$ results using (16$\times$16$\times$2) k-points, providing converged results.

Most importantly, we observe that the $GW$ band structure also exhibit a direct band gap at the $\Gamma$ point. LDA yields a bandgap at the $\Gamma$ point of 0.23 eV, thereby underestimating the gap by $0.38$ eV compared to the GW results. When correcting for the LDA band gap by applying a scissors shift of 0.38 eV to the LDA conduction bands, the overall agreement with the bands trend in LDA and GW is good, although there are small discrepancies in higher lying conduction bands and deeper lying valance bands. These differences between the LDA eigenvalues and quasi-particle energies also depend on the $k$-points. For example, at the high symmetry points the difference is up to $0.1$ eV. However, to a reasonable approximation, the quasi-particle eigenvalues can be obtained from the Kohn-Sham eigenvalues by rigidly shifting the conduction bands. In particular, one has to keep in mind that the overall accuracy of $GW$ calculations is in the order of 0.1 eV only. Thus, it can be argued that for the solution of the BSE and the corresponding dielectric function, also Kohn-Sham results based on LDA eigenvalues can be used, just incorporating the scissors operator to circumvent the heavy $GW$ calculations for all $k$-points. Nevertheless, we have also undertaken calculations in which we also take into account the full quasi-particle band structure as input for the solution of the BSE to calculate the optical transitions.
\subsection{BSE Results}

\begin{figure}[!htbp]
	\includegraphics[width=\columnwidth]{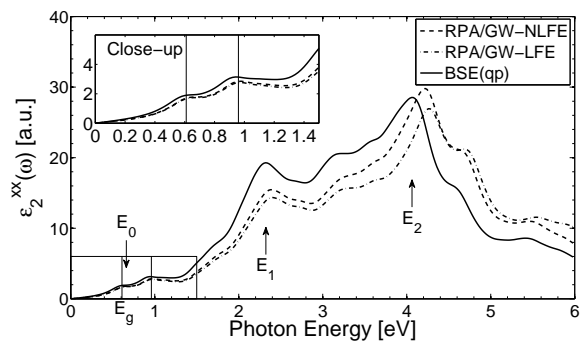}
	\caption{\label{xx_RPA_BSE}Imaginary part of the in-plane ($xx$) component of the dielectric function of $\alpha_{12}$ calculated within RPA without LFEs(dashed line) and with LFEs(dash-dot line). The BSE (solid line) based on quasi-particle band-structure at the $G_0W_0$ level. Two vertical lines are added at peak position of first two transitions in order to better see the red-shift in BSE results. The inset magnifies the onset region of the spectrum. }
\end{figure}

Fig.~\ref{xx_RPA_BSE} shows the imaginary part of the dielectric function $\epsilon_2(\omega)$ for in-plane polarization (electric field polarized along $\hat x$) obtained by two different approximations. The dashed line represents RPA results which ignore both electron-hole correlations and also local field effects(LFEs). The dash-dot line shows the effects of local field on RPA results. The BSE(qp) result (solid line) corresponds to solving the BSE for the dielectric function on top of a $G_0W_0$ quasi-particle band structure.
The most prominent effect of including excitonic effects, \emph{i.e.} when going from the RPA to BSE result, are changes in the strength of two main $E_1$ and $E_2$ transitions. Note that the direct, attractive term of the BSE-kernel dominates over the repulsive exchange term (responsible for the local field effects). The repulsive exchange term results in slightly blue shift of the transitions, and also weaker oscillation strength, a finding which is in accordance with results for bulk Si and Ge.\cite{Albrecht1998,Benedict98-2}  While  $E_1$ is enhanced by approximately one third, the oscillator strength of $E_2$ is slightly diminished and its peak position is shifted to smaller energies. A similar observation has also been made for bulk Si and Ge for which BSE calculations were shown to improve the agreement with experiment.\cite{Albrecht1998,Benedict98-2,Rohlfing00,Arnaud2001,Puschnig2002a}

The peak of the onset transition ($E_0$) appears at $0.61$ eV, which corresponds to the quasi-particle bandgap at the $\Gamma$ point reported in the previous section. This feature originates from a transition of the VBM (this band is labeled later on 40) and the CBM (band number 41). We have also evaluated the second prominent transition at $0.94$ eV by looking into  the dipole matrix elements for several band combinations. This leads us to conclude that it is most-likely arising from a transition of band 40 to band 43, also at the $\Gamma$ point. This transition corresponds to the 1.05 eV absorption line reported in Ref.~\onlinecite{Avezac2012}.
When interpreting these onset transitions, one has to point out, however, that a correct calculation of the optical properties in this small energy range probably would require the inclusion of phonons into the Hamiltonian which is beyond the scope of the present study. Moreover, the intrinsic accuracy of the $GW$+BSE approach, in particular as a result of the critical $k$ point sampling, is only in the order of 0.1 eV. This is also illustrated by the fact that the lowest energy transition is shifted to slightly larger energies when a symmetry-breaking $k$-grid is used as is depicted in Fig.~\ref{so_bse_gw} of the Appendix. Here, the $E_0$ peak appears at $0.75$ eV, and  also the next absorption band lies at the somewhat larger energy of 1.16eV. Due to the finite lifetime broadening of $0.15$ eV necessary to efficiently carry out the Haydock recursion scheme, it is not possible to distinguish additional absorption lines between $0.61$ eV and $1.16$ eV.

Nevertheless we can conclude that the inclusion of electron-hole interactions also enhances the oscillator strengths of the two low-energy direct transitions ($E_0$) and the one at 0.94 eV. When going from RPA to BSE, there are also small red shifts of the transition energies visible that are, however, only about $0.03$ eV in size. Such small shifts are below the overall accuracy  expected for the set of computational parameters chosen in this work, and more precise calculations are necessary in future to evaluate the excitonic nature of these transitions further.

\begin{figure}[!htbp]
	\includegraphics[width=\columnwidth]{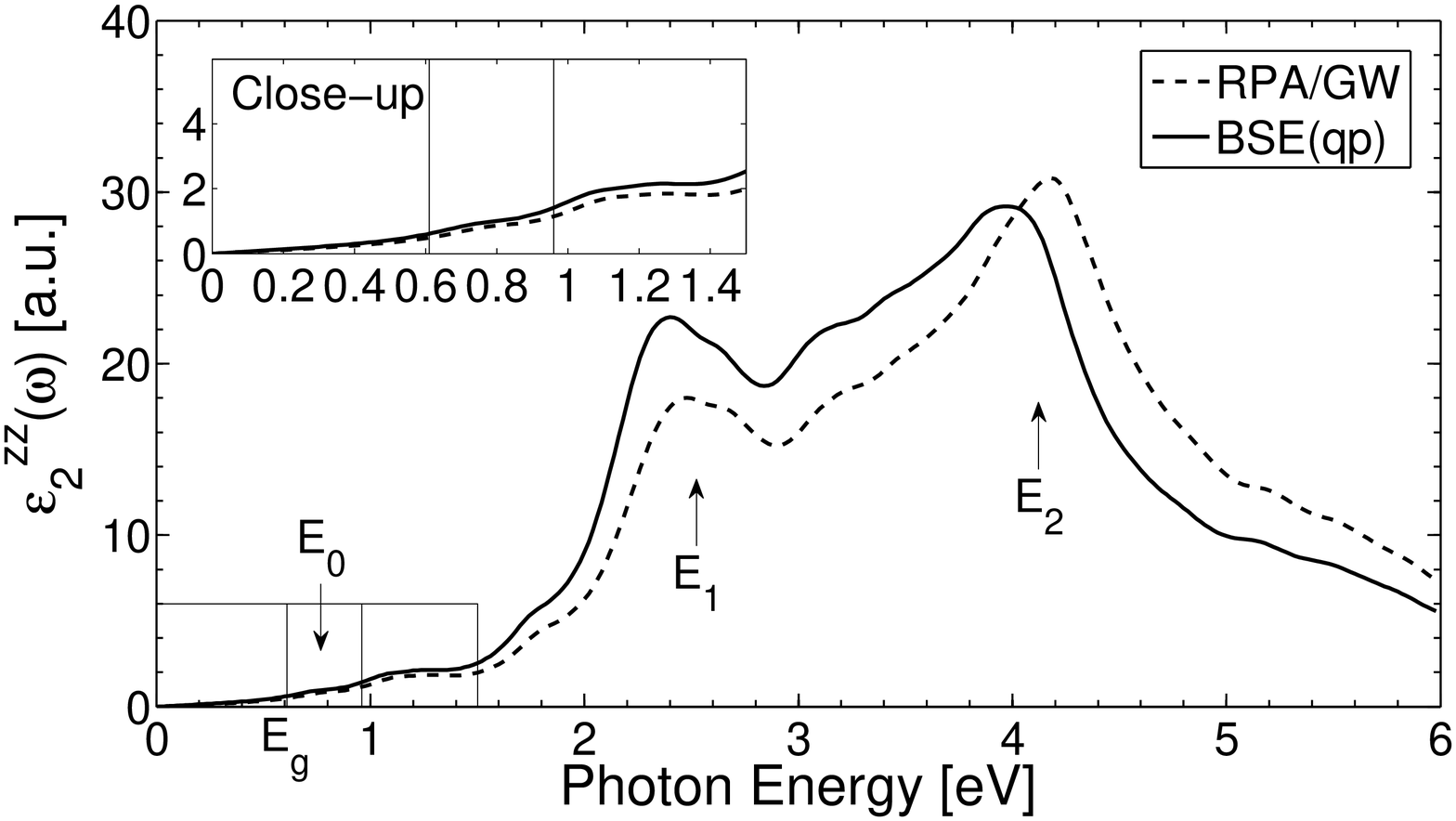}
	\caption{\label{zz_RPA_BSE}Imaginary part of the out-of-plane ($zz$) component of the dielectric function of $\alpha_{12}$ calculated within in RPA (dashed line) and BSE (solid line) based on quasi-particle band-structure at the $G_0W_0$ level. Two vertical lines are added at peak position of first two transitions in order to better see the red-shift in BSE results.  The inset magnifies the onset region of the spectrum.}
\end{figure}

We now turn to the anisotropy of the dielectric tensor.  Fig.~\ref{zz_RPA_BSE} shows RPA and BSE(qp) results for the out-of-plane polarization.  As already discussed in the d’'Avezac's paper for the $\alpha_{12}$ structure, the in-plane ($xx$) imaginary part of the dielectric function at low energies is higher than the out-of-plane $\epsilon_2$ which is also visible from Fig.~\ref{zz_RPA_BSE}. The oscillator strengths of the onset transitions is suppressed compared to the $xx$ polarization. This also implies that the onset of photoluminescence transitions, at $0.61$ eV, will be TE (s-) polarized. 

\begin{figure}[!htbp]
	\includegraphics[width=\columnwidth]{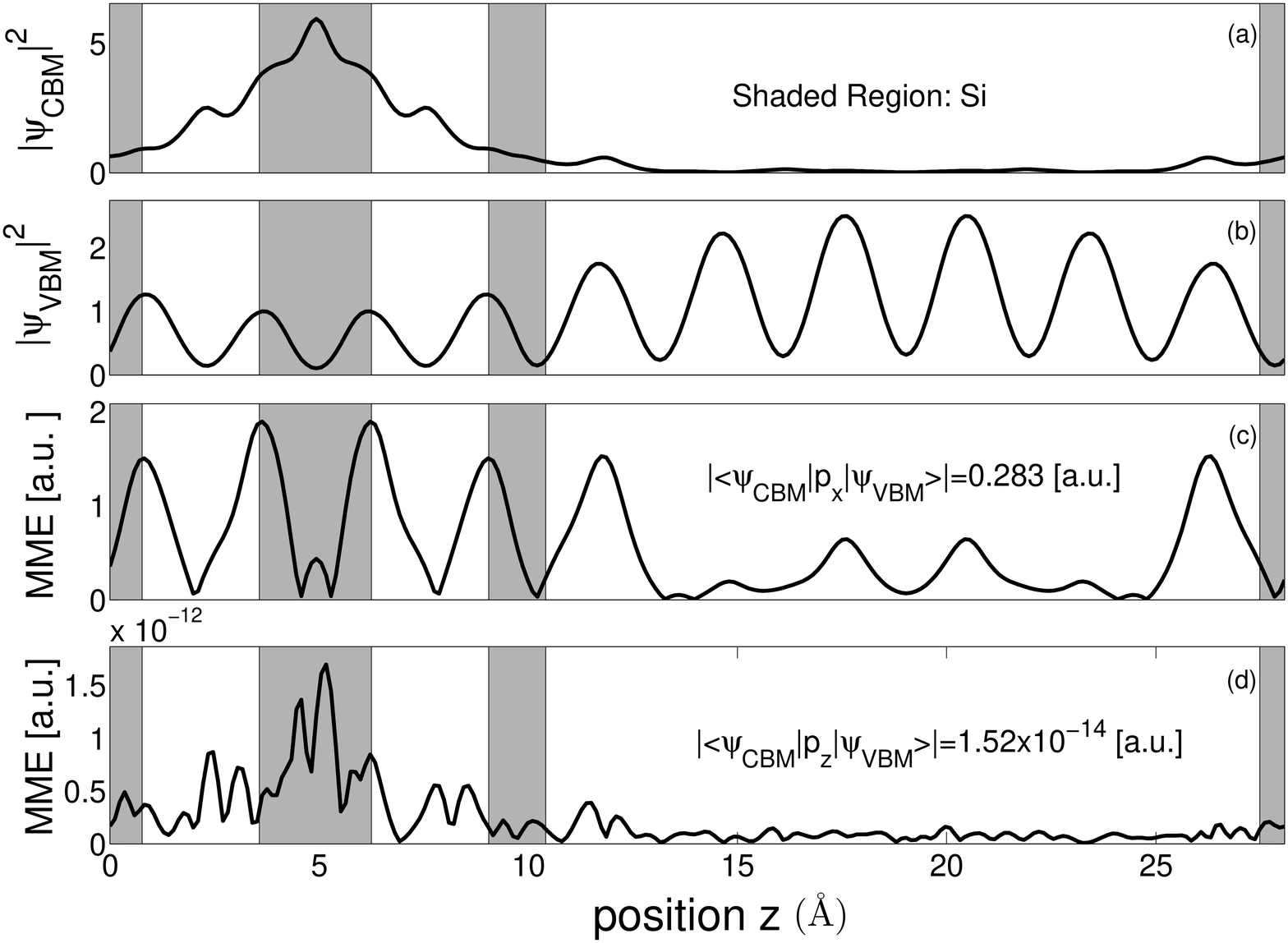}
	\caption{\label{mme}Panels (a) and (b) show the modulus square of the CBM and VBM wave functions, respectively, averaged in the $xy$ plane. The lower two panels depict the $z$-dependent momentum matrix elements plotted along growth direction as defined in the text for in-plane (c) and out-of-plane (d) polarization at the $\Gamma$ point for the onset transition. }
%	\label{mme}
\end{figure}

In order to understand the origin for this anisotropy further and to reveal possible reasons for the presence of a direct transitions at the $\Gamma$ point, we investigate the respective transition matrix elements. To this end, we plot the square of the modulus of VBM and CBM wavefunctions averaged over the $xy$ plane in panels (a) and (b) of Fig.~\ref{mme} for VBM and CBM, respectively. In addition to this analysis which has already been used in Ref.~\onlinecite{Avezac2012}, we also spatially decompose the transition matrix elements. So, we define the quantities $p_{xx}(z)$ and $p_{zz}(z)$ , in the following way,

\begin{equation}
{p_{xx}}(z) = \iint{\,dx \,dy {\psi _{CBM}^*({x},{y},{z}){{\mathrm{p}}_{x}}{\psi_{VBM}}({x},{y},{z)}}}
\end{equation}
\begin{equation}
{p_{zz}}(z) = \iint{\,dx \,dy {\psi _{CBM}^*({x},{y},{z}){{\mathrm{p}}_{z}}{\psi_{VBM}}({x},{y},{z)}}}
\end{equation}

in order to decompose the  $z$-dependent contributions to the transition matrix elements averaged over the epitaxial plane $(x,y)$. Fig.~\ref{mme} shows the absolute value  of the momentum matrix elements (MME) calculated in electric dipole approximation, where the wave vector of electromagnetic wave is considered much smaller than the size of Brillouin zone. The matrix elements can be obtained for all k-points, but the main interest is in the polarization dependent (in-plane and out-of-plane) transition from the VBM  to the CBM.

It has been argued before that the optical transition arises mainly from the Si motif region (shaded region). However, when calculating the $z$-dependent and in plane polarized transition matrix element we see that at four out of the six Si-Ge interfaces the absolute value of the transition matrix element shows a maximum and at the two others, it is small. This is similar to previous finding that surface optical spectra are mainly due to ``extrinsic'' contributions directly related to the surface chemistry, rather than ``intrinsic'' effects due to contribution within the bulk.\cite{Hahn2001,Aspnes1985} We expect that this insight can be employed for a further optimization of the direct emission properties of Si-Ge structures. For a better comparison, we also show in panels (a) and (b) the modulus square of the CBM and VBM wave functions, respectively, averaged in the $xy$ plane. It is worth noting that the in plane optical transitions originate also from the Ge buffer layer despite the fact that the electrons in the CBM are mainly located in the Si motif region.

 The out-of-plane matrix element is  basically zero, so it is expected that photoluminescence is TE polarized. This also shows that the absolute values of matrix elements are slightly different for the onset transition at $\Gamma$ point than the values found in Ref.~\onlinecite{Avezac2012}. This could be due to the use of different pseudo-potentials or a slightly different strain state.

%%%%%%%%%%%%%%%%%%%%%%%%%%%%%%%%%%%%%%%%%%%% Conclusion %%%%%%%%%%%%%%%%%%%%%%%%%%%%%%%%%%% 
\section{Conclusion}

In this paper we used many-body perturbation theory to calculate the quasi-particle energies and optical properties of SiGe$_2$Si$_2$Ge$_2$SiGe$_{12} $ (or $\alpha_{12}$) structure. $G_0W_0$ calculations for the band structure corrects the band gap compared to LDA results. To good approximation, the $G_0W_0$ band structure can be understood by a $k$-independent upward shift of the LDA conduction states (scissors shift), where small deviations in the order of 0.1 eV become visible for higher lying conduction states. Optical properties including excitonic effects are calculated on top of a quasi-particle band structure by solving the Bethe-Salpeter equation. We find two distinct transitions at 0.61 eV and 0.94 eV arising from the weakly-allowed direct transition at the $\Gamma$ point. Although excitonic effects for these onset transition are not strong, we observe the typical enhancement of oscillator strengths also seen for the higher lying $E_1$ transition band in Si and Ge bulk structures accompanied by an overall weak red shift of the adsorption features. However, it must be noted that the overall accuracy of our present $GW$+BSE approach does not allow for a quantitative analysis of the exciton binding energy of this onset transition. By a spatial decomposition of optical transition matrix elements, we found that the onset transitions are mainly arising from Si-Ge interface with additional contributions from Ge buffer layer. This admits the idea that surface optical spectra is mainly due to ``extrinsic'' contributions, rather than ``intrinsic'' effects due to contribution within the bulk.
  
%%%%%%%%%%%%%%%%%%%%%%%%%%%%%%%%%%%%%%%%%%%  Acknowledgement %%%%%%%%%%%%%%%%%%%%%%%%%%%%%%
\section*{Acknowledgement}
We gratefully thank for all the support of Johann Messner from the University Linz supercomputer department. We acknowledge financial support from the Austrian Science Fund (FWF) project P23190-N16.

%%%%%%%%%%%%%%%%%%%%%%%%%%%%%%%%%%%%%%%%%%%% Appendix %%%%%%%%%%%%%%%%%%%%%%%%%%%%%%%%%%%%%
\section{Appendix}
\label{appendix}
\appendix

In this appendix, we include several convergence tests which highlight the dependence of our results with respect to a few critical computational parameters. We focus firstly on the influence of the plane wave cut-off used for representing the response functions, secondly on the number of bands included in the transition space for setting up the BSE Hamiltonian, and thirdly we discuss the convergence of optical spectra with $k$-point sampling. Due to the strong dependence of computational time and memory demands on these parameters, such careful tests are important to assess the overall accuracy of the presented results.

\begin{figure}[!htbp]
	\includegraphics[width=0.7\columnwidth]{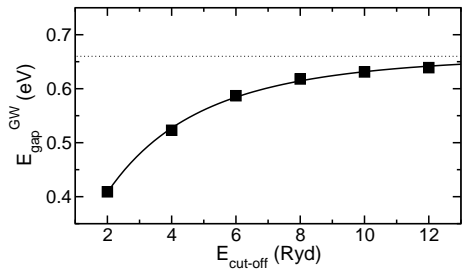}
	\caption{\label{gap_ecuteps}Convergence of the $G_0W_0$ gap at the $\Gamma$ point as function of the plane wave cut-off for the response function computed for a moderate $k$ mesh of $6 \times 6 \times 1$. At the chosen value of 6 Rydberg, an error (underestimation) of the band gap in the order of 0.1 eV may be expected.}
\end{figure}

\begin{figure}[!htbp]
	\includegraphics[width=0.7\columnwidth]{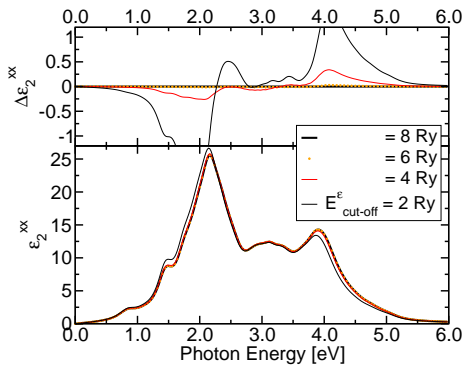}
	\caption{\label{bse_ecuteps}Convergence of BSE spectra with the plane wave cut-off for the response function computed for a moderate $k$ mesh of $6 \times 6 \times 1$. The bottom panel shows spectra based on a LDA band structure with a fixed scissors shift for four different cut-off values. In the top panel, the differences to the converged spectrum computed for an 8 Rydberg cut-off are shown. }
\end{figure}

In a plane wave code, the response functions, \emph{i.e.}~the polarizability and the screened Coulomb interaction, are represented in a plane wave basis as well. Usually one can work with much smaller cut-off values as compared to wave functions, nevertheless, a careful choice is important due to the unfavorable scaling of computational time and memory demands with this parameter. It affects both, the calculation of the quasi-particle band structure as well as the calculation of optical spectra through the screened Coulomb interaction matrix elements entering the BSE Hamiltonian. Fig.~\ref{gap_ecuteps} shows how the quasi-particle gap at the $\Gamma$ point depends on this cut-off energy. Note that for the sake of computational time, this convergence test was performed with a rather small $k$-point sampling of $6 \times 6 \times 1$. In order to reach convergence an energy cut-off of 12 Rydberg seems enough, however, already a modest value of 6 Rydberg is expected to yield a quasi-particle gap with an accuracy somewhat better than 0.1 eV. When solving the BSE based on LDA eigenvalues corrected by a fixed scissors shift for the conduction bands for various plane wave cut-off values in the screened Coulomb interaction, we observe a very fast convergence as can be seen from Fig.~\ref{bse_ecuteps}. Already at 6 Rydberg, the computed spectrum are converged and basically  indistinguishable from the 8 Rydberg result.

\begin{figure}[!htbp]
	\includegraphics[width=\columnwidth]{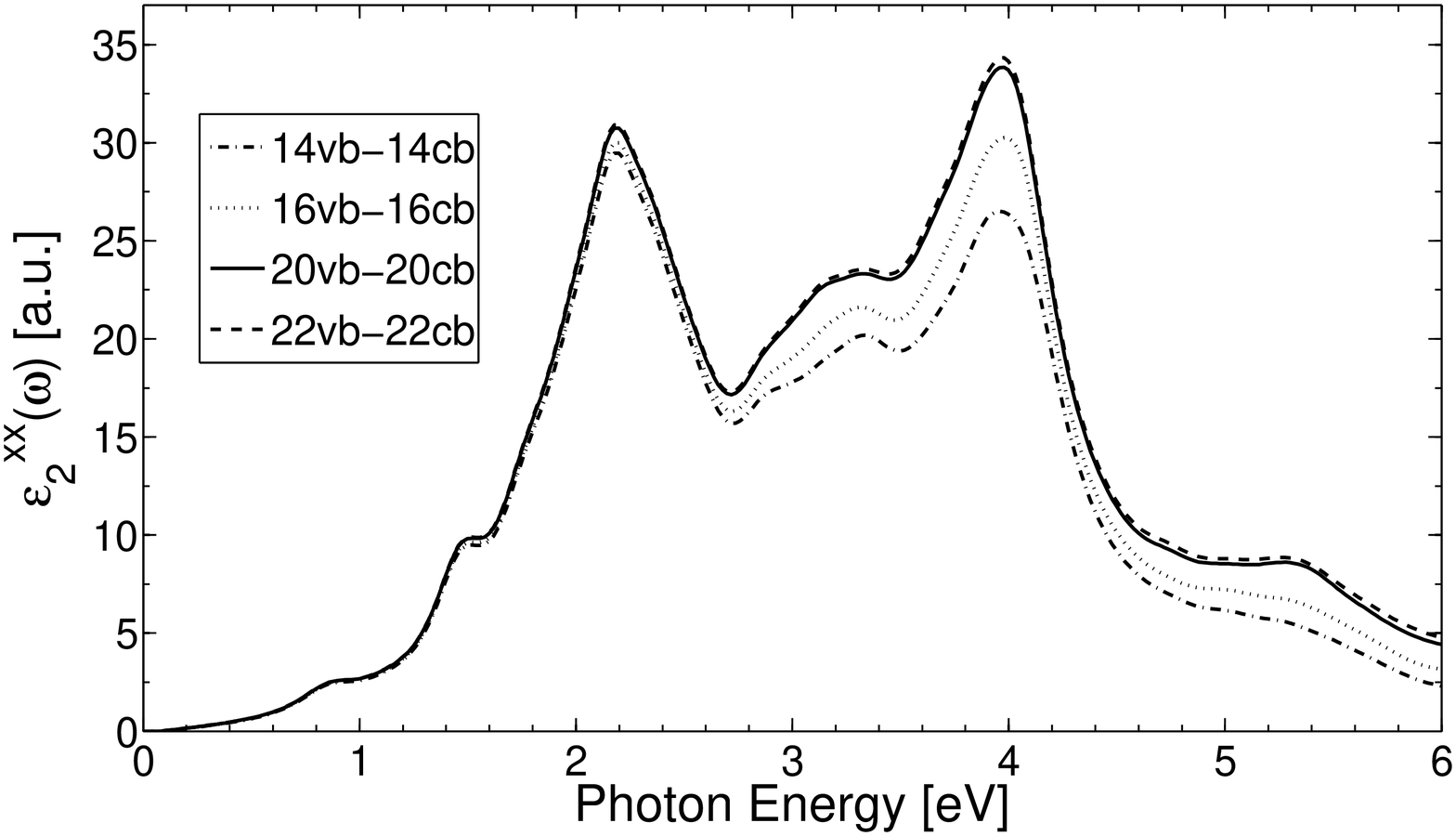}
	\caption{\label{conv_bands}Convergence test for imaginary part of dielectric function, calculated by solving BSE (s.o. method), in which the number of bands included in the transition space has been varied.}
\end{figure}

When determining the eigenvalues and -vectors of the BSE Hamiltonian of Eq.~(\ref{Hamiltonian}), the transition space has to be truncated. In Fig.~\ref{conv_bands}, the convergence of imaginary part of the dielectric function in the range of 0-6 eV with respect to the number of valence and conduction states included in the BSE Hamiltonian is displayed. Note that for this test a moderate $k$ point grid of 6$\times$6$\times$1 has been used. We conclude that a total of 40 bands, \emph{i.e.}~20 valence and 20 conduction bands leads to converged spectra, in particular, in the low energy range of the spectrum, which is of primary interest in this study, while at somewhat higher transition energies larger deviations become visible.

\begin{figure}[!htbp]
	\includegraphics[width=\columnwidth]{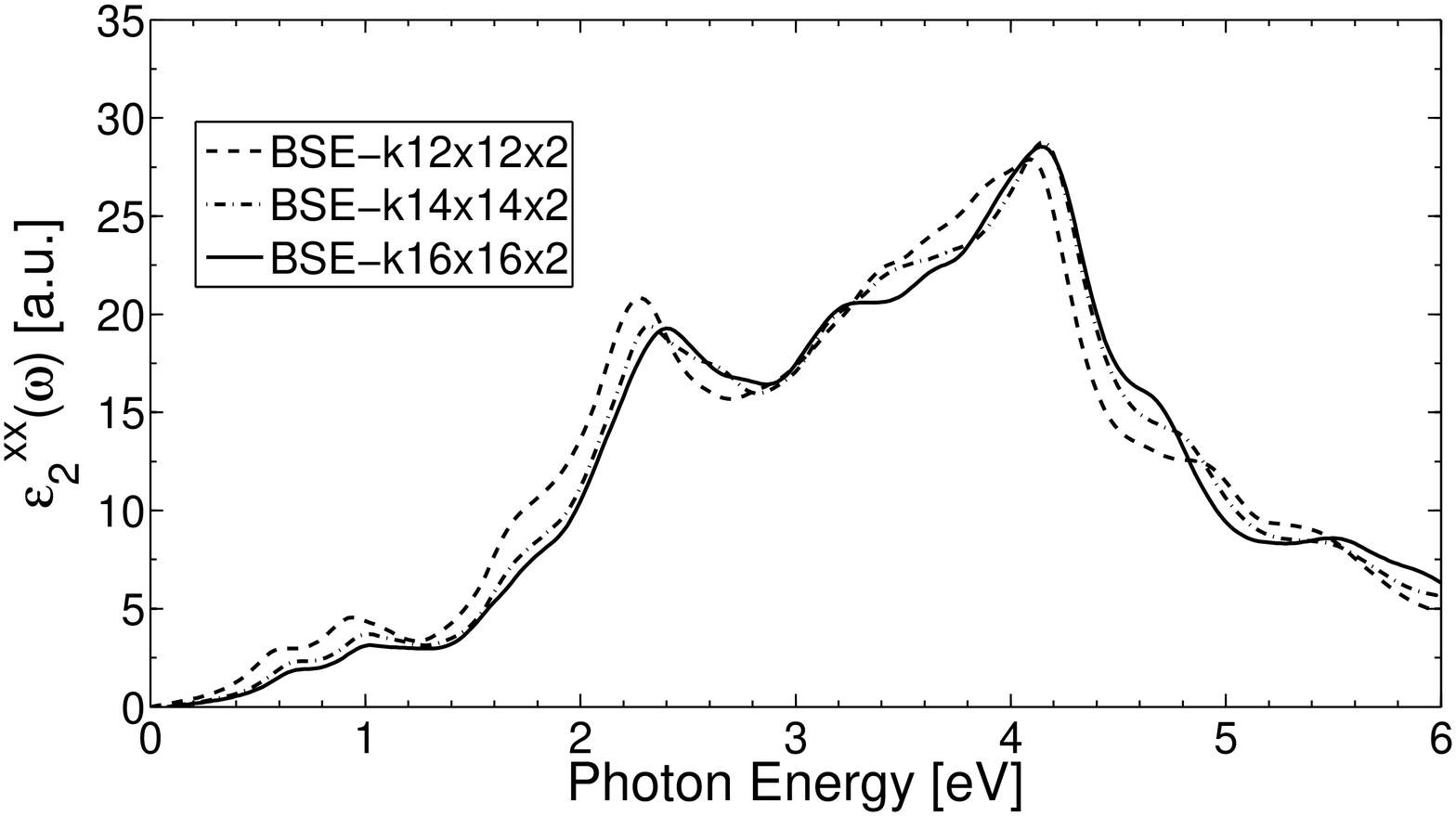}
	\caption{\label{epsilon2_qp_energy}Convergence test for imaginary part of dielectric function, calculated by solving BSE (qp method), over number of k-points in BZ. }
\end{figure}

\begin{figure}[!htbp]
	\includegraphics[width=\columnwidth]{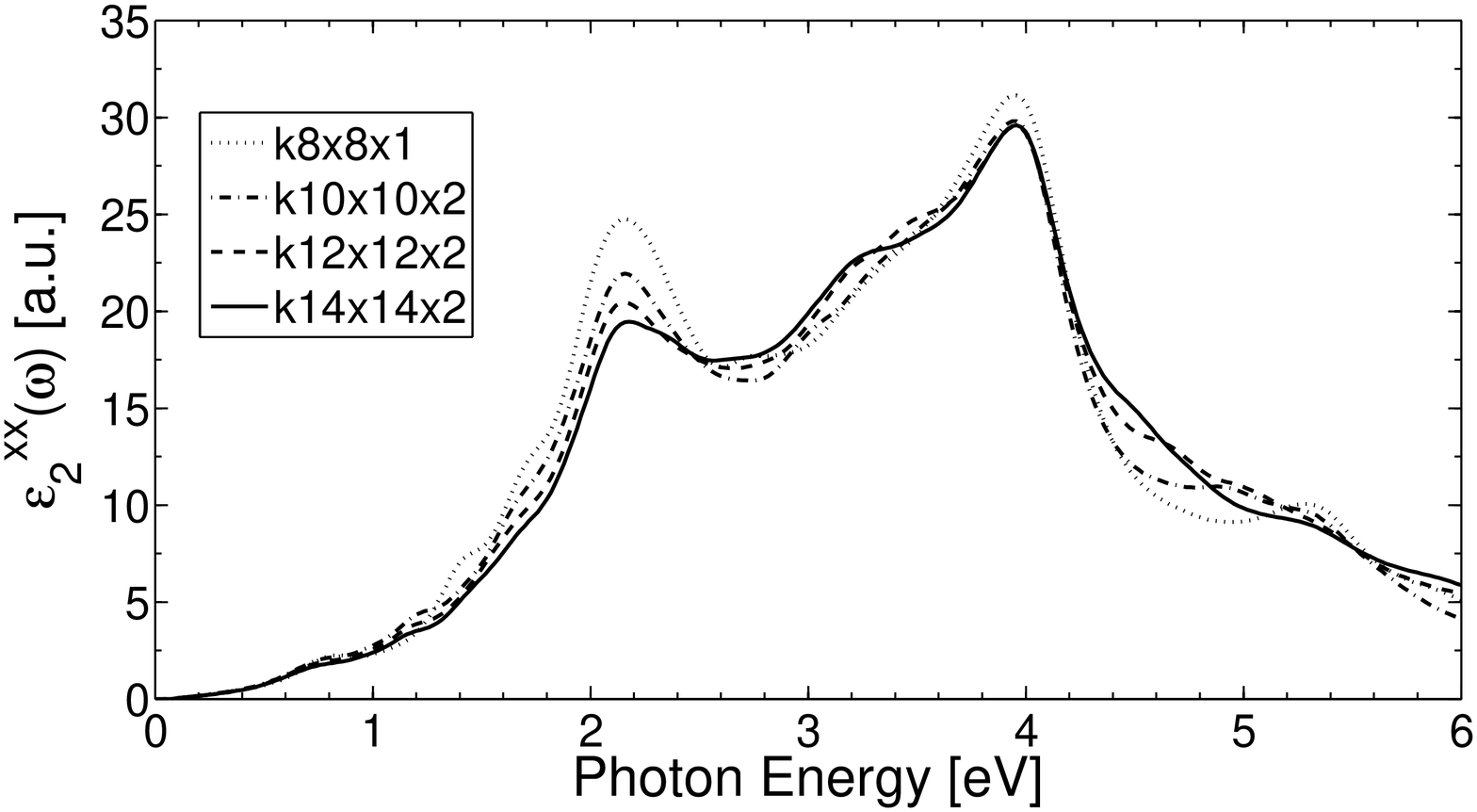}
	\caption{\label{conv_kpoints}Convergence test for imaginary part of dielectric function, calculated by solving BSE (scissors operator method), over number of k-points in BZ.}
\end{figure}

A very critical computational parameter for obtaining converged optical spectra is the $k$ point sampling of the Brillouin zone (BZ). Since in common implementations of the BSE, the full BZ has to be sampled, it critically influences the size of the resulting BSE Hamiltonian, and thus the memory and CPU demands. 
Fig.~\ref{epsilon2_qp_energy} shows a convergence test for the in-plane $\epsilon_2(\omega)$  obtained from the solution of the BSE for various $k$ point grids. Note that these results are based on the quasi-particle band structure obtained from a preceding $G_0W_0$ calculation and utilize a $\Gamma$-centered $k$ point grid. We observe that the spectrum at the most dense grid, $16 \times 16 \times 2$ can be expected to be reasonably converged in terms of transition energies and oscillator strengths of the main $E_1$ and $E_2$ transition bands. The peak heights of the low-energy onset transitions are, however, not fully converged at this $k$ mesh.

\begin{figure}[!htbp]
	\includegraphics[width=\columnwidth]{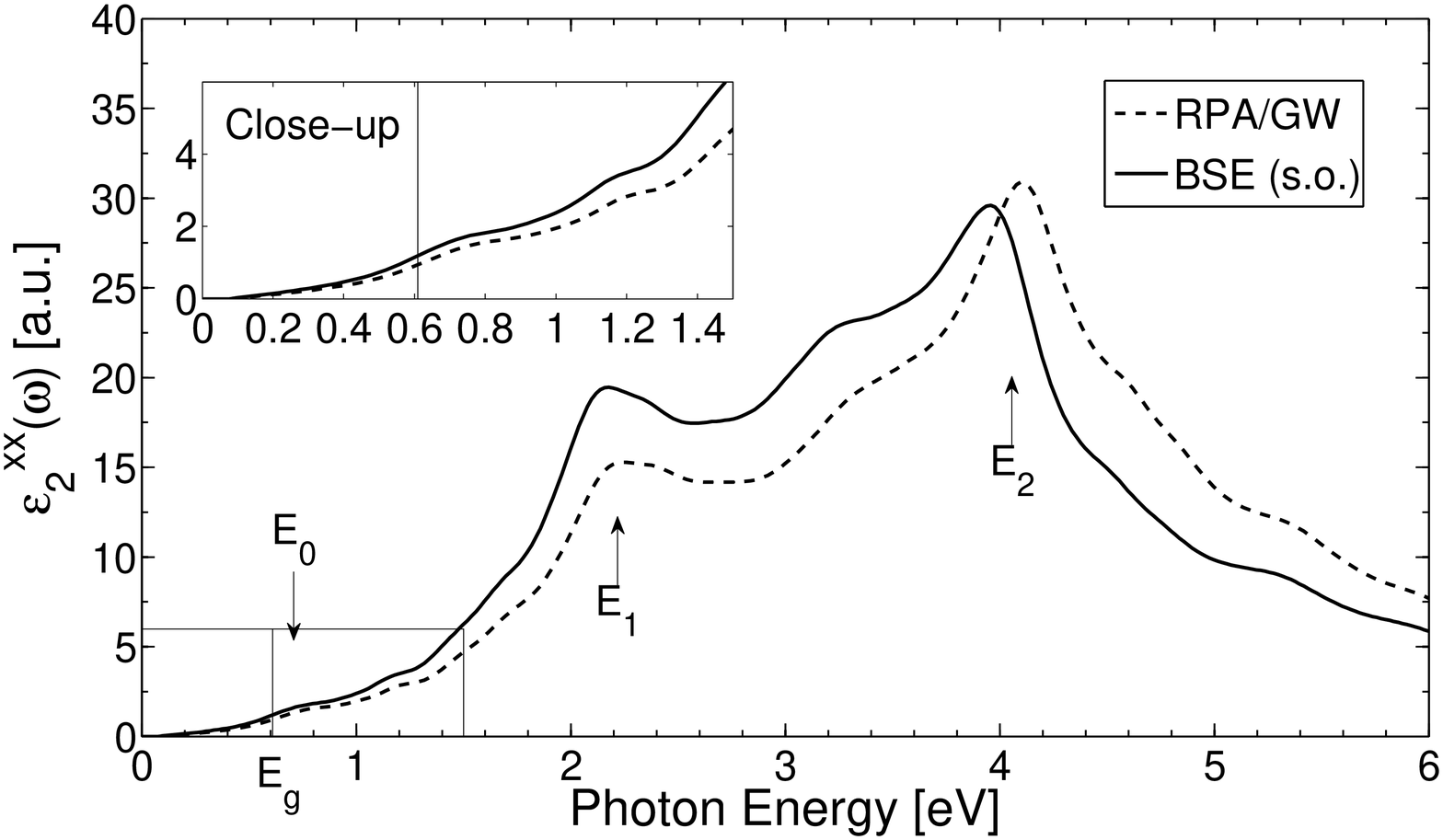}
	\caption{\label{so_bse_gw}Imaginary part of the in-plane ($xx$) dielectric function of $\alpha_{12}$ calculated within RPA (dashed line) and BSE (solid line) based on a LDA band structure corrected by a scissors operator shift for the conduction bands. Note that a shifted $k$ mesh of $14 \times 14 \times 2$ has been used.}
\end{figure}

In order to analyze the influence of the $k$-point sampling further, we have also utilized a $k$-point grids shifted by  with respect to the $\Gamma$-point, in particular a symmetry-breaking shift of ($0.11$, $0.21$, $0.31$) has been used. Results for such shifted $k$ meshes for various grid densities are displayed in Fig.~\ref{conv_kpoints}. The appearance of the main $E_1$ and $E_2$ transition bands and their convergence with respect to the number of $k$ points is similar to the case of the non-shifted $k$-grid. The low-energy onset transitions originating from the direct band gap at the $\Gamma$ point, however, turn out to be less pronounced and also shifted to slightly larger transition energies. This can be seen in Fig.~\ref{so_bse_gw} that compares RPA with BSE results and emphasize the onset transition in its inset. Clearly, the weak onset set transitions appears slightly shifted to higher energies compared to the position of the direct, quasi-particle gap indicated by the vertical line. The reason is that the shifted $k$ does no longer include the direct gap at $\Gamma$ which leads to a shift of this direct transition to higher energies. For that reason, we have chosen to discuss results originating from the $\Gamma$-centered grids in the main part of the paper even though a symmetry-breaking grid of the same grid density might exhibit a slightly better level of convergence.

We conclude this appendix about various convergence issues by noting that with the current settings, the $G_0W_0$+BSE calculations for the $\alpha_{12}$ structure composed of 20 atoms of silicon and germanium are highly demanding, both in terms of CPU time and memory usage. The presented results are  thus a compromise limited by the available hardware capacities (Altix UV 1000 shared memory machine with 360 CPUs and $\approx$2 TBytes memory). However, by providing the full convergence tests in this appendix, the achieved level of computational accuracy can be assessed. 
%

%\bibliography{list}

%Merlin.mbs v4.21 2009-07-09.
%

\end{document}